\documentclass[prb,twocolumn,floatfix,preprintnumbers,superscriptaddress,amsmath,amssymb]{revtex4}

\usepackage{graphicx}
\usepackage{dcolumn}
\usepackage{bm}

\begin{document}

\title{Density of states and supercurrent in diffusive SNS junctions:
role of nonideal interfaces and spin-flip scattering}

\author{J.C. Hammer}
\affiliation{Fachbereich Physik, Universit\"at Konstanz, D-78457 Konstanz, 
Germany}
\affiliation{Institut f\"ur Theoretische Festk\"orperphysik,
Universit\"at Karlsruhe, D-76128 Karlsruhe, Germany}

\author{J.C. Cuevas}
\affiliation{Departamento de F\'{\i}sica Te\'orica de la Materia
Condensada, Universidad Aut\'onoma de Madrid, E-28049 Madrid, Spain}
\affiliation{Institut f\"ur Theoretische Festk\"orperphysik,
Universit\"at Karlsruhe, D-76128 Karlsruhe, Germany}
\affiliation{Forschungszentrum Karlsruhe, Institut f\"ur Nanotechnologie, 
D-76021 Karlsruhe, Germany}

\author{F.S. Bergeret}
\affiliation{Departamento de F\'{\i}sica Te\'orica de la Materia
Condensada, Universidad Aut\'onoma de Madrid, E-28049 Madrid, Spain}

\author{W. Belzig}
\affiliation{Fachbereich Physik, Universit\"at Konstanz, D-78457 Konstanz, 
Germany}

\begin{abstract}
We present a theoretical study of the density of states and supercurrent in
diffusive superconductor-normal metal-superconductor (SNS) junctions. In particular,
we study the influence on these two equilibrium properties of both an arbitrary 
transparency of the SN interfaces and the presence of spin-flip scattering in
the normal wire. We show that the minigap that is present in the spectrum of the 
diffusive wire is very sensitive to the interface transmission. More importantly,
we show that at arbitrary transparency the minigap replaces the Thouless energy as 
the relevant energy scale for the proximity effect, determining for instance the 
temperature dependence of the critical current. We also study in detail
how the critical current is suppressed by the effect of spin-flip scattering,
which can be due to either magnetic impurities or, under certain circumstances,
to an external magnetic field. Our analysis based on the quasiclassical
theory of diffusive superconductors can be very valuable to establish 
quantitative comparisons between experiment and theory. 
\end{abstract}

\maketitle

\section{Introduction}

When a normal metal (N) and a superconductor (S) are brought together, their
mutual interaction results in the modification of their electronic and transport 
properties. In particular, the normal metal may acquire genuine superconducting
properties such a gap in the density of states or the ability to sustain
a supercurrent. This effect, known as \emph{proximity effect}, was first discussed
by de Gennes~\cite{deGennes1964} in the 1960's and in recent years it has been 
extensively studied in diffusive hybrid nanostructures.~\cite{Pannetier2000}
Many equilibrium~\cite{Gueron1996,Moussy2001} and transport 
properties~\cite{Courtois1999,Dubos2001} of diffusive SN systems are now well 
understood, which is partially due to the impressive predictive power of the 
quasiclassical theory of superconductivity for diffusive systems, which is 
summarized in the Usadel equations.~\cite{Usadel1970}

The proximity effect is mediated by Andreev reflections.~\cite{Andreev1964}
In this tunneling process an electron coming from N with energy $\epsilon$ below 
the superconducting gap $\Delta$ is converted into a reflected hole, thus 
transferring a Cooper pair to the S electrode. The time-reversed states involved 
in this process are coherent over a distance $L_C = \mbox{min} (\sqrt{\hbar D 
/\epsilon}, L_{\phi})$, where $D$ is the diffusion constant of N and $L_{\phi}$ 
is the phase coherence length. This coherence may also be altered by interactions 
that break the time-reversal symmetry such as those induced by paramagnetic 
impurities or an external magnetic field.

In this work we present a theoretical study of the density of states and the 
supercurent in diffusive SNS junctions. These quantities nicely reflect the proximity 
effect under equilibrium conditions.  It was first shown by McMillan~\cite{McMillan1968} 
that a diffusive normal metal in contact with a superconductor can develop a gap
in its electronic spectrum, which is usually referred to as \emph{minigap}.
More recently the minigap has been studied by numerous authors, usually
within the framework of the Usadel equations.~\cite{Golubov1988,Belzig1996,
Zhou1998,Ivanov2002,Crouzy2005} From the experimental point of view, the
appearance of a minigap has been tested indirectly in several tunneling
experiments (see for instance Refs.~[\onlinecite{Scheer2001,Rubio-Bollinger2003}]
an references therein).

On the other hand, the fact that a SNS junction can sustain a supercurrent
is known since the first experiments performed with Pb-Cu-Pb 
sandwiches.~\cite{Clarke1969,Shepherd1972} It was soon realized that the
existence of a dissipationless current in these structures is possible due to
the proximity effect.~\cite{deGennes1964} Later on, a more systematic experimental
study of the critical current in these hybrid structures was carried out
with the help of diffusive SNS microbridges.~\cite{Warlaumont1979,Dover1981}
The results of these experiments were described by Likharev,~\cite{Likharev1976}
who made use of the Usadel equations in the high temperature regime ($\Delta \ll 
k_{\rm B}T$). A more general study of the Josephson effect in diffusive SNS 
junctions was made in Ref.~[\onlinecite{Zaikin1981}]. More recently, Dubos
\emph{et al.}~\cite{Dubos2001} demonstrated that the full temperature 
dependence of the critical current of diffusive Nb-Cu-Nb junctions with 
highly-transparent interfaces could be quantitatively described by the 
quasiclassical theory.

Most of the theoretical work done on proximity effect has been focused in 
the case of either ideal (perfectly transmissive) SN interfaces or in the tunneling
limit,~\cite{Kupriyanov1988,Volkov1993,Golubov2004} with some notable 
exceptions.~\cite{Heikkila2002} One of the two main goals of this paper is to 
study how the local density of states (DOS) and the supercurrent in diffusive SNS 
junctions is influenced by arbitrary transmission of the interfaces. This is an 
important issue, in particular, in order to be able to establish quantitative 
comparisons between theory an experiment, since in reality the mismacht of material 
parameters leads to a broad range of transmission through the SN
interfaces. In particular, we shall discuss the following issues:
(i) how the transmission determines the magnitude of the minigap and, in
general, the shape of the DOS in the normal wire, both in the absence and 
in the presence of a supercurrent and (ii) how a finite transmission 
modifies the current-phase relation and the critical current of these
junctions. Our results, based on the quasiclassical theory, show that
the minigap, which is reduced as the interface resistance increases,
is the energy scale that controls, in particular, the magnitude and 
temperature dependence of the critical current. For ideal interfaces,
this role is played by the Thouless energy $\epsilon_T = \hbar D / L^2$, 
where $L$ is the length of the normal metal.

The second goal of our work is to study the role of spin-flip scattering
in the local DOS and supercurrent of diffusive SNS junctions. This type
of scattering, which can be induced by magnetic impurities or an external 
magnetic field, breaks the time reserval symmetry between the electrons
in Cooper pairs and reduces the superconducting correlations.~\cite{Abrikosov1965,
Maki1966} Different authors~\cite{Volkov1993,Belzig1996,Yip1995,Yokoyama2005,
Crouzy2005} have studied the effect of spin-flip scattering in the properties 
of SN structures. However, basic quantities like the supercurrent in SNS
structures have, to our knowledge, not yet been addressed. We present in this 
work a detailed study of the dependence of the critical current of a diffusive
SNS on the strength of the spin-flip scattering for arbitrary range of
parameters (length of the normal wire, temperature and interface resistance).
Our predictions can tested experimentally by measuring the critical current
in the presence of magnetic field since, as long as the normal wire is a thin 
film, the field acts simply as a pair-breaking mechanism.~\cite{Maki1966,
Belzig1996,Anthore2003} 

The rest of the paper is organized as follows. In the next section we describe
the general formalism, based on the quasiclassical theory for diffusive
superconductors, that allows us to compute the DOS and the supercurrent in
diffusive SNS junctions for arbitrary length, temperature and interface
transparency. Section III is devoted to the analysis of the local DOS in
the normal wire in different situations. In section IV we discuss the
results for the dependence of the supercurrent on the interface resistance, 
temperature and strength of the spin-flip scattering. Finally, we summarize
our main conclusions in section V. In the Appendix A we describe our 
analytical results for the critical current in the limit of weak proximity
effect and in Appendix B we include some numerical fits described in the
previous sections.

\section{Quasiclassical Green's functions formalism}

\begin{figure}[t]
\begin{center}
\includegraphics[width=0.75\columnwidth,clip]{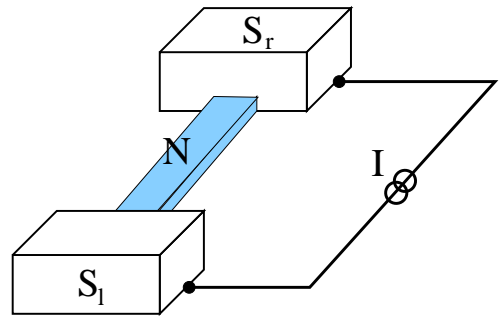}
\caption{\label{setup} (Color online) Schematic representation of the system:
a metallic diffusive wire (N) is connected at its ends to superconducting reservoirs
S$_l$ and S$_r$. Eventually, a supercurrent $I$ may circulate along the
SNS junction.}
\end{center}
\end{figure}

We consider the SNS junction represented schematically in Fig.~\ref{setup}, 
where N is a diffusive normal metal of length $L$ coupled to two identical 
superconducting reservoirs with gap $\Delta$. We assume that the transport
is phase-coherent, i.e. $L \ll L_{\phi}$ and neglect the suppression of the
pair potential in the S leads near the interfaces. Our main goal is to study 
how the equilibrium properties of this system are influenced by the transparency
of the SN interfaces and by the presence of a spin-flip mechanism in the
diffusive wire. In particular, we want to study (i) the equilibrium density of
states (DOS) in the normal wire and (ii) the supercurrent in the SNS system when
a superconducting phase difference $\phi$ is established between the electrodes.

In order to describe these properties we use the quasiclassical theory of 
superconductivity in the diffusive limit,~\cite{Usadel1970,Larkin1986,Belzig1999}
where the mean free path is much smaller than the superconducting coherence length
in the normal metal, $\xi=\sqrt{\hbar D/ \Delta}$. This theory is 
formulated in terms of momentum averaged Green's function ${\bf \check G}({\bf R},\epsilon)$, 
which depend on position ${\bf R}$ and an energy argument $\epsilon$, since we shall only deal 
with stationary situations. This propagator is a $4\times 4$ matrix in Keldysh space 
(indicated by an inverted caret), where each entry is a $2\times 2$ matrix in electron-hole 
space (indicated by a caret)
\begin{equation}
\label{keldysh-space}
{\bf \check G} = \left( \begin{array}{cc}
\hat G^R & \hat G^K \\
   0     & \hat G^A
\end{array} \right); \hspace{5mm} 
\hat G^{R} = \left( \begin{array}{cc}
{\cal G}^{R} & {\cal F}^{R} \\
\tilde {\cal F}^{R}  & \tilde {\cal G}^{R}
   \end{array} \right) .
\end{equation}
The general definitions of the different functions can be found in 
Ref.~[\onlinecite{Serene1983}]. The Green's functions for the left (l) and right (r) 
leads can be written as ${\bf \check G}_{j}(\epsilon) =
e^{-i \phi_j \hat \tau_3/2\hbar} {\bf \check G}_0(\epsilon)
e^{i \phi_j \hat \tau_3/2 \hbar}$, where $\phi_j$ is the phase
of the order parameter of the electrode $j=l,r$. Here, ${\bf \check G}_0(\epsilon)$ 
is the equilibrium bulk Green's function of a BCS superconductor.
Notice that, since we shall only consider equilibrium situations, the Keldysh
component of ${\bf \check G}({\bf R},\epsilon)$ can be expressed in terms of
the retarded and advanced components as $\hat G^K = (\hat G^R - \hat G^A) 
\tanh(\beta \epsilon /2)$, where $\beta = 1/k_{\rm B} T$ is the inverse of the 
temperature.

The propagator ${\bf \check G}({\bf R},\epsilon)$ satisfies the 
stationary Usadel equation, which in the N region reads \cite{interaction}
\begin{equation}
\label{usadel-eq}
 \frac{\hbar D}{\pi} \nabla \left( {\bf \check G} \nabla {\bf \check G} 
 \right) - \frac{\hbar}{2\pi \tau_{sf}} \left[ {\bf \check \tau_3} {\bf \check G}
{\bf \check \tau_3} , {\bf \check G} \right] + 
\epsilon \left[ {\bf \check \tau_3},  {\bf \check G} \right] = 0 ,
\end{equation}
\noindent
where ${\bf \check \tau_3}$ is proportional to the unit matrix in Keldysh space
and equal to the Pauli matrix $\hat{\tau}_3$ in electron-hole space. Equation
(\ref{usadel-eq}) is supplemented by the normalization condition ${\bf \check G}^2
= -\pi^2 {\bf \check 1}$. In the previous equation, $\tau_{sf}$ is the scattering 
time associated to spin-flip (magnetic) impurities or related pair-breaking mechanisms.
For instance, as it has been shown in Refs.~[\onlinecite{Belzig1996,Anthore2003}], if the normal
wire is a thin film and its width $W$ does not exceed $\xi$, the effect of a perpendicular 
magnetic field $H$ can be described with an effective spin-flip scattering rate
$\Gamma_{sf} = \hbar/ \tau_{sf} = D e^2 H^2 W^2/(6 \hbar)$.

In order to solve numerically the Usadel equation it is convenient to use the so-called Riccati
parametrization,\cite{Eschrig2000} which accounts automatically for the normalization
condition. In this method and for spin-singlet superconductors, the retarded and
advanced Green's functions are parametrized in terms of two coherent functions 
$\gamma^{R,A}({\bf R},\epsilon)$ and $\tilde \gamma^{R,A}({\bf R}, \epsilon)$ 
as follows
\begin{eqnarray}
\label{riccati}
\hat G^{R,A} & = & \mp i \pi 
\hat N^{R,A} \left( \begin{array}{cc}
1 - \gamma^{R,A} \tilde \gamma^{R,A} & 2 \gamma^{R,A} \\
2 \tilde \gamma^{R,A} & \tilde \gamma^{R,A} \gamma^{R,A} -1
\end{array} \right) ,
\end{eqnarray}
with the ``normalization matrices"
\begin{equation}
\hat N^{R,A} = \left(
\begin{array}{cc}
(1 + \gamma^{R,A} \tilde \gamma^{R,A})^{-1} & 0 \\
0 & (1 + \tilde \gamma^{R,A} \gamma^{R,A} )^{-1}
\end{array} \right) \nonumber .
\end{equation}

Some of these functions are related by fundamental symmetries (particle-hole,
retarded-advanced) like
\begin{eqnarray}
\gamma^A({\bf R},\epsilon)  =  
- \left[ \tilde \gamma^R( {\bf R},\epsilon) \right]^* &; & 
\gamma^A({\bf R},\epsilon) = - \gamma^R( {\bf R},-\epsilon) .
\end{eqnarray}
Therefore, we just have to determine, for instance, the retarded functions.
Using their definition in Eq.~(\ref{riccati}) and the Usadel equation 
(\ref{usadel-eq}), one can obtain the following transport equations for 
these functions in the normal wire region~\cite{Eschrig2004}
\begin{eqnarray}
\partial^2_x \gamma^R + (\partial_x \gamma^R) \frac{\tilde {\cal F}^R}{i\pi}
 (\partial_x \gamma^R) - 2 \left( \frac{\Gamma_{sf}}{\epsilon_T} \right) 
\gamma^R \frac{\tilde {\cal G}^R}{i\pi}
& & \nonumber \\
\label{g-eq}
+ 2i \left( \frac{\epsilon}{\epsilon_T} \right) \gamma^R & = & 0 \\
\partial^2_x \tilde \gamma^R + (\partial_x \tilde \gamma^R) \frac{{\cal F}^R}{i\pi}
 (\partial_x \tilde \gamma^R) + 2 \left( \frac{\Gamma_{sf}}{\epsilon_T} \right)
\tilde \gamma^R \frac{{\cal G}^R}{i\pi}
& & \nonumber \\
\label{gtilde-eq}
+ 2i \left( \frac{\epsilon}{\epsilon_T} \right) \tilde \gamma^R & = & 0 .
\end{eqnarray}
\noindent
Here, $x$ is the dimensionless coordinate which describes the position along 
the N wire and ranges from 0 (left lead) to 1 (right lead). The expressions for 
$\tilde {\cal F}^R$, $\tilde {\cal G}^R$, ${\cal F}^R$ and ${\cal G}^R$ are 
obtained by comparing Eq.~(\ref{keldysh-space}) with Eq.~(\ref{riccati}).
Notice that Eqs.~(\ref{g-eq},\ref{gtilde-eq}) couple the functions with and without
tilde. This means in practice that, in general, one has to solve 
Eqs.~(\ref{g-eq},\ref{gtilde-eq}) simultaneously.

Now, we have to provide the boundary conditions for the Eqs.~(\ref{g-eq},\ref{gtilde-eq}).
Let us first remind that for ideal interfaces (perfect transparency) such 
conditions at the ends of the N wire result from the continuity of the Green's 
functions over the SN interfaces:
\begin{eqnarray}
\label{ideal}
 \gamma^R_l (\epsilon) = \gamma^R_0(\epsilon) &;&
\tilde \gamma^R_l (\epsilon) = - \gamma^R_0(\epsilon) \nonumber \\
 \gamma^R_r (\epsilon) = e^{-i \phi} \gamma^R_0(\epsilon-eV) &;&
 \tilde \gamma^R_r (\epsilon) = - e^{i \phi} \gamma^R_0(\epsilon+eV) ,
\end{eqnarray}
where $\gamma^R_l(\epsilon) \equiv \gamma^R (x=0,\epsilon)$ and 
$\gamma^R_r(\epsilon) \equiv \gamma^R (x=1,\epsilon)$, and the same
for the coherent function with tilde. Here, $\gamma^R_0(\epsilon ) = 
-\Delta /\{\epsilon^R + i\sqrt{\Delta^2-(\epsilon^R)^2} \}$, where
$\epsilon^R = \epsilon+i0^+$. Finally, $\phi$ is the eventual phase 
difference between the two superconducting reservoirs, which we assume 
to be applied in the right electrode.

For non-ideal interfaces one has to use the more general boundary conditions
derived in Refs.~[\onlinecite{Nazarov1999,Kopu2004}]. These conditions for an 
spin-conserving interface are expressed in terms of the Green's functions as follows
\begin{equation}
{\bf \check G}^{\beta} \partial_x {\bf \check G}^{\beta} = \left( \frac{G_0}
{G_{\rm N}} \right) \sum_i \frac{2\pi^2 \tau_i \left[ {\bf \check G}^{\beta}, 
{\bf \check G}^{\alpha} \right]} {4\pi^2 - \tau_i \left( 
\left\{{\bf \check G}^{\beta}, {\bf \check G}^{\alpha} \right\} + 2\pi^2
\right)} .
\end{equation}
Here, ${\bf \check G}^{\beta(\alpha)}$ refers to the Keldysh-Green's function
on side $\beta(\alpha)$ of the interface, $G_0=2e^2/h$ is the quantum of 
conductance, $G_{\rm N}$ is the conductance of the normal wire and $\tau_i$ are the
different transmission coefficients characterizing the interface. In general,
one would need the whole set $\{ \tau_i \}$, but since one does not
have access to this information we adopt here a practical point of view.
We assume that all the $N$ interface open channels have the same
transmission $\tau$ and define $G_B = G_0 N \tau$ as the conductance of the barrier.
Thus, the two S-N interfaces will be characterized by two quantities, namely
the barrier conductance $G_{\rm B}$ and the transmission $\tau$, and our starting
point for the boundary conditions will be
\begin{equation}
\label{non-ideal}
r {\bf \check G}^{\beta} \partial_z {\bf \check G}^{\beta} = 
\frac{2\pi^2 \left[ {\bf \check G}^{\beta},
{\bf \check G}^{\alpha} \right]} {4\pi^2 - \tau \left(
\left\{{\bf \check G}^{\beta}, {\bf \check G}^{\alpha} \right\} + 2\pi^2
\right)} ,
\end{equation}
where we have defined the ratio $r=G_{\rm N}/G_{\rm B}$. In this language, an ideal interface
is characterized by $r=0$ and a tunnel contact is described by $\tau \ll 1$. In 
what follows, unless the opposite is explicitly stated, we shall assume a symmetric 
situation with two identical interfaces. In the literature the so-called 
Kupriyanov-Lukichev~\cite{Kupriyanov1988} boundary conditions are often used. 
These conditions can be obtained from Eq.~(\ref{non-ideal}) by removing the term 
proportional to $\tau$ in the denominator. Such approximation is valid strictly 
speaking for the case of tunnel junctions ($\tau \ll 1$) and it turns out to be 
very good for highly transparent interfaces ($r \ll 1$).

The next step is to express these boundary conditions directly in terms of the 
coherent functions. Substituting the definitions of 
Eq.~(\ref{riccati}) into Eq.~(\ref{non-ideal}) and after straightforward algebra, 
one obtains the following boundary conditions for the parameterizing functions
\begin{eqnarray}
\label{g-boundary}
\mp r \frac{\partial_x \gamma^R_{\beta} + (\gamma^R_{\beta})^2 \partial_x 
\tilde \gamma^R_{\beta}} {(1 + \gamma^R_{\beta} \tilde \gamma^R_{\beta})^2}
= &  & \nonumber \\
 \frac{(1 - \gamma^R_{\beta} \tilde \gamma^R_{\beta}) \gamma^R_{\alpha}
- (1 - \gamma^R_{\alpha} \tilde \gamma^R_{\alpha}) \gamma^R_{\beta}}
{(1 + \gamma^R_{\beta} \tilde \gamma^R_{\beta}) (1 + \gamma^R_{\alpha} 
\tilde \gamma^R_{\alpha}) - \tau (\gamma^R_{\alpha} - \gamma^R_{\beta})
(\tilde \gamma^R_{\alpha} - \tilde \gamma^R_{\beta})} &  & ,
\end{eqnarray}
where the minus sign is for the left interface and the plus sign for the right one.
The boundary conditions for $\tilde \gamma^R$ can be obtained from Eq.~(\ref{g-boundary})
exchanging the quantities without tilde by the corresponding ones with tilde 
and vice versa. These equations establish a relation between the functions and their derivatives
evaluated on the side of the interface inside the N wire ($\beta$) and the corresponding
functions evaluated on the side of the interface inside the reservoir ($\alpha$),
which are given by Eq.~(\ref{ideal}).

In the limit of weak proximity effect, Eqs.~(\ref{g-eq},\ref{gtilde-eq}) can 
solved analytically, as we discuss in Appendix A. However, in general
they have to be solved numerically. These
are second order differential equations with boundary conditions relating the
functions and their derivatives in the two SN interfaces. This is a typical 
two point boundary value problem that we solve numerically using the so-called 
relaxation method as described in Ref.~[\onlinecite{relaxation}]. We want to
point out that the Riccati parametrization facilitates the numerical calculations
because the coherent functions are smooth and bounded. Moreover, this
parametrization is also well suited for  time-dependent
problems, as we have shown in Ref.~[\onlinecite{Cuevas2006}].

To end this section we discuss the formula for the supercurrent. The electrical current 
can expressed in terms of the Usadel Green's functions as~\cite{Larkin1986}
\begin{equation}
I = \frac{G_{\rm N}}{8\pi^2 e} \int^{\infty}_{-\infty} d\epsilon \; \mbox{Tr}
\left\{ \hat \tau_3 \left[ {\bf \check G} \partial_x {\bf \check G}
\right]^K (\epsilon ) \right\}.
\end{equation}
Combining this expression with fundamental symmetries of the Green's functions 
and using the fact that we only address equilibrium situations, we can write 
the supercurrent as
\begin{equation}
\label{superc}
I = \frac{G_{\rm N}}{e} \int^{\infty}_{-\infty} d\epsilon \; \mathcal S(\epsilon) 
\tanh (\beta \epsilon /2) ,
\end{equation}
\noindent
where $\mathcal S = (1/4 \pi^2) \mbox{Re} \{ \mbox{Tr} ( \hat \tau_3 
\hat G^R \partial_x \hat G^R) \}$ is the spectral supercurrent.

\section{Local density of states}

In equilibrium, the most basic quantity that reflects the proximity effect in the N wire is
the local density of states (DOS), which is defined as $\mbox{DOS}(x,\epsilon) =
- \mbox{Im} \left\{ {\cal G}^{R}(x,\epsilon)  \right\} /\pi$. This quantity can in
principle be measured with a tunneling probe electrode as in Ref.~[\onlinecite{Gueron1996}],
or with a scanning tunneling microscope (STM) as in Ref.~[\onlinecite{Moussy2001}].
In this section we analyze the local DOS in the normal wire in different situations.

\begin{figure}[!t]
\begin{center}
\includegraphics[width=0.85\columnwidth,clip]{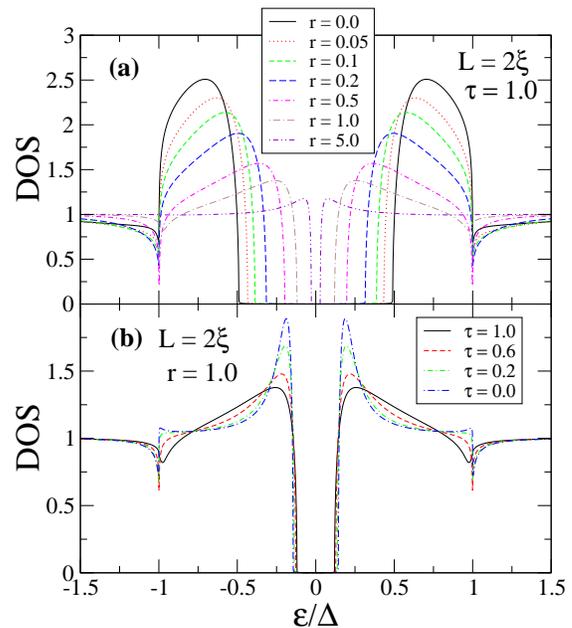}
\caption{\label{dos-sns} (Color online) Density of states of a SNS junction as a
function of energy in the middle of a wire of length $L=2\xi$ without spin-flip
scattering ($\Gamma_{sf} = 0$). The SN interfaces are assumed to be identical
and there is no phase difference between the S electrodes. In panel (a) the
different curves correspond to different values of the ratio $r=G_{\rm N}/G_{\rm B}$ 
and a transmission $\tau=1$, while in panel (b) they correspond to different
values of the transmission $\tau$ for a ratio $r=1.0$.}
\end{center}
\end{figure}

Let us start discussing a situation where there is no phase difference between the
superconducting reservoirs ($\phi=0$). In this case, one can show that the relation
$\tilde \gamma^R(\epsilon) = - \gamma^R(\epsilon)$ holds. Thus, one only needs to
solve Eq.~(\ref{g-eq}) for the coherent function $\gamma^R(\epsilon)$. 
In Fig.~\ref{dos-sns} we show an example of the local DOS in the middle of
a normal wire ($x=0.5$) of length $L=2\xi$ without spin-flip scattering 
($\Gamma_{sf} = 0$). The most prominent feature is the appearance of 
a minigap $\Delta_g$ in the spectrum, which for perfect transparency scales 
with the Thouless energy roughly as $\Delta_g \sim 3.1 \epsilon_T$ in the
long junction limit ($L \gg \xi$). Let us remind that the minigap is the same
along the normal wire, although the exact DOS depends on the position. The 
existence of a minigap in a diffusive normal metal in contact with a superconductor 
was discussed by McMillan~\cite{McMillan1968} within a tunneling model, where the normal 
region was a thin layer. In more recent years, the minigap has been extensively 
studied in various hybrid diffusive SN and SNS structures.~\cite{Golubov1988,
Belzig1996,Zhou1998} As one can see in Fig.~\ref{dos-sns}(a), this minigap 
diminishes progressively as the ratio $r$ increases, i.e. as the interface 
becomes more opaque. For this particular length, we find that the minigap 
decays with the interface parameter $r$ as $\Delta_g/\Delta \sim 0.14/r$ for 
$r > 1$ (see the fit to our numerical data in Fig.~\ref{fit}, Appendix B).

In Fig.~\ref{dos-sns}(b) we illustrate the effect of the transmission coefficient
$\tau$ in the local DOS for a ratio $r=1.0$. Notice that the minigap is only
slightly reduced as $\tau$ decreases, while the features around $\Delta_g$ become 
more pronounced. The effect of a transmission smaller than one is much more 
pronounced for larger values of $r$, i.e $r \gg 1$, while for values $r < 1$ it is
rather insensitive to the value of $\tau$.

\begin{figure}[t]
\begin{center}
\includegraphics[width=0.85\columnwidth,clip]{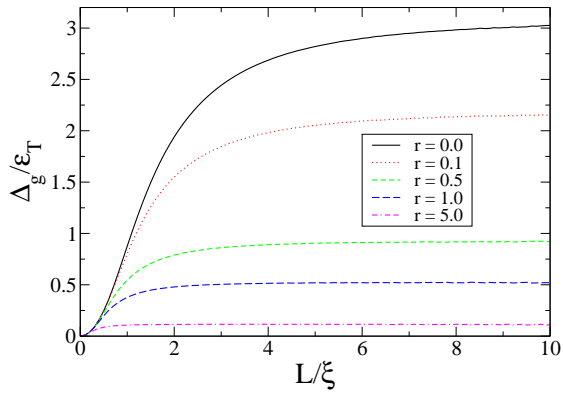}
\caption{\label{minigap} (Color online) Minigap $\Delta_g$ of a SNS junction as a 
function of the length of the N wire for different values of $r=G_{\rm N}/G_{\rm B}$ 
and $\Gamma_{sf} = 0$. The contact is assumed to be symmetric and the transmission 
is set to $\tau=1$. Notice that is $\Delta_g$ is normalized by the Thouless energy
$\epsilon_T$.}
\end{center}
\end{figure}

In Fig.~\ref{minigap} we present a detailed study of the decay of the minigap
as a function of the wire length for different values of the interface resistance
and $\tau=1.0$. We have normalized the minigap $\Delta_g$ with the Thouless energy
to show explicitly that in the long wire limit $\Delta_g$ simply scales with this
energy. In this limit ($\Delta/\epsilon_T \rightarrow \infty$) we were able to
fit accurately the decay of the minigap with the ratio $r$ with the function
$\Delta_g/ \epsilon_T = 0.64 / (0.20 + r)$ (see Fig.~\ref{fit} in Appendix B).
In the opposite case of a short junction, i.e. when $L\ll \xi$, the minigap is 
of the order of $\Delta$ for perfect transparency, while it is given by $\Delta_g
\sim \epsilon_T/2r$ in the limit of $r\gg 1$.~\cite{Volkov1993}

\begin{figure}[t]
\begin{center}
\includegraphics[width=0.9\columnwidth,clip]{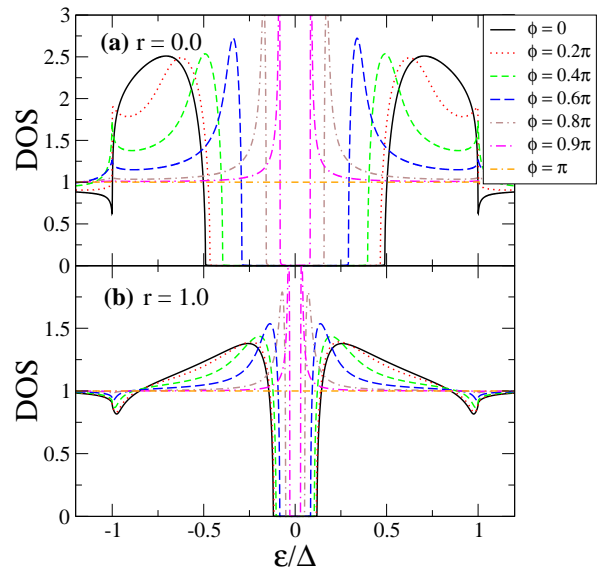}
\caption{\label{dos-phi} (Color online) Density of states of a SNS junction as a
function of energy in the middle of a wire of length $L=2\xi$ ($\Gamma_{sf} = 0$)
for different values of the superconducting phase difference $\phi$. The interfaces 
are identical and characterized by a ratio $r=G_{\rm N}/G_{\rm B}= 0.0$ in panel (a) and 
$r=1.0$ in panel (b). In both cases $\tau=1$.}
\end{center}
\end{figure}

Let us now study how the density of states is modified when there is a finite
phase difference $\phi$ between the leads, i.e. in the presence of a supercurrent.
In this discussion we shall assume that $\Gamma_{sf} = 0$. Considering ideal interfaces,
Zhou \emph{et al.}~\cite{Zhou1998} showed theoretically that the minigap decreases
monotonically as the phase difference increases and it closes completely when $\phi =
\pi$. In Fig.~\ref{dos-phi} we show two examples for $L=2\xi$ of how 
the DOS in the middle of the wire evolves with the phase $\phi$ for perfect 
transparency and $r=1.0$. Notice that for finite $r$ the qualitative behavior of
the minigap is very similar. Indeed, a detailed study shows that, if the
minigap is normalized by its value at $\phi = 0$, its phase dependence
does not change significantly with the interface resistance. Notice, however,
that the features in the DOS around the minigap can be clearly different, as
Fig.~\ref{dos-phi} exemplifies. 

Now we turn to the analysis of the influence of spin-flip scattering in the
local DOS. Belzig \emph{et al.}~\cite{Belzig1996} showed that the 
minigap of an SN structure is reduced in the presence of a spin-flip mechanism
and vanishes for large values of $\Gamma_{sf}$. Different authors~\cite{Volkov1993,
Yip1995,Yokoyama2005} have studied the effect of magnetic impurities in the transport
of SN structures and found that the Thouless energy is the scale that controls the 
effect of spin-flip on the proximity effect. In particular, Crouzy \emph{et 
al.}~\cite{Crouzy2005} have shown analytically that in the long junction limit
of an SNS structure, the minigap closes at a critical value of $\Gamma^C_{sf} 
\approx 4.96 \epsilon_T$. 

\begin{figure}[t]
\begin{center}
\includegraphics[width=0.9\columnwidth,clip]{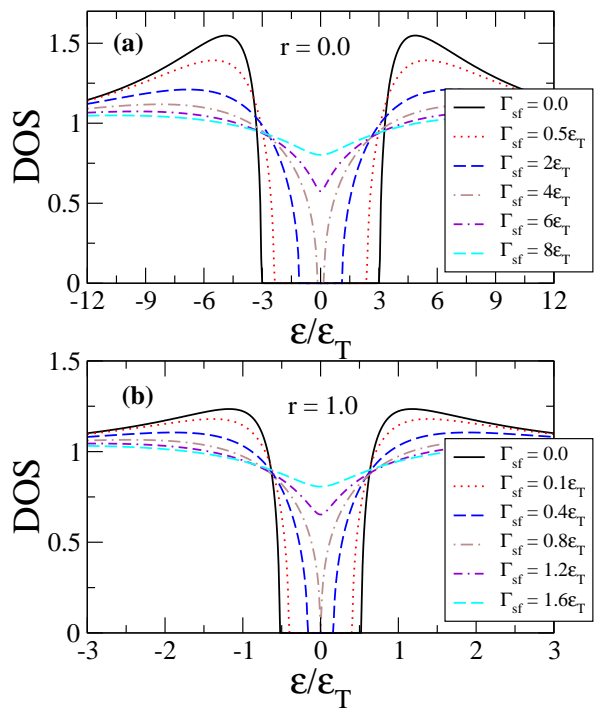}
\caption{\label{dos-sf} (Color online) (a) Density of states of a SNS junction as a
function of energy in the middle of a wire of length $L=10\xi$ for $r=0$ and
$\tau=1.0$. The different curves correspond to different values of the spin-flip
scattering rate $\Gamma_{sf}$. (b) The same but for $r=1.0$. Notice that both
the energy and the $\Gamma_{sf}$ are in units of the Thouless energy.}
\end{center}
\end{figure}

Fig.~\ref{dos-sf} displays the local DOS in the middle of a normal wire of
length $L=10\xi$ ($\epsilon_T = 0.01\Delta$) for different values of the 
spin-flip rate $\Gamma_{sf}$. The upper panel shows the case of ideal interfaces,
while the lower one contains the results for a ratio $r=1.0$. One can see how
the minigap is progressively reduced as $\Gamma_{sf}$ increases and finally
vanishes. For $r=0$ (perfect interfaces) we find numerically that the gap closes
at $\Gamma^C_{sf} \approx 4.9 \epsilon_T$ in very good agreement with the long 
junction limit mentioned above.~\cite{Crouzy2005} For the case $r=1.0$ this 
critical value is $\Gamma^C_{sf} \approx 0.8 \epsilon_T$. This means that it is 
reduced by approximately a factor 6, which is the same reduction factor obtained
for the minigap (see Fig.~\ref{minigap}). This indicates that at finite 
transmission the relevant scale for the proximity effect is the minigap
rather than the Thouless energy. This will become even clearer in the 
analysis of the supercurrent in the next section.

\section{Supercurrent}

As mentioned in the introduction, the supercurent in diffusive SNS junctions
has been the subject of numerous theoretical and experimental studies. In
particular, from the theory side, the results for the critical current 
for ideal interfaces and without spin-flip scattering are summarized in
Refs.~[\onlinecite{Dubos2001,Golubov2004}]. The critical current in SNS junctions 
with partially transparent interfaces was discussed in 
Ref.~[\onlinecite{Kupriyanov1988}] using the boundary conditions developed
in the same reference. More recently, Heikkil\"a \emph{et al.}~\cite{Heikkila2002}
studied the reduction of the zero-temperature critical current with the interface 
resistance considering a disordered interface.

In this section we shall discuss how both the supercurrent and the critical
current are modified by a finite transparency of the interfaces. To be precise,
we shall investigate both the current-phase relationship and the 
temperature dependence of the critical current. Moreover, we shall study in 
detail the effect of spin-flip scattering in the critical current, which to our
knowledge has not been discussed before in the literature. This analysis is
very relevant from the experimental point of view, since it might describe the
supercurrent in the presence of a magnetic field, as explain in the previous 
section. Finally, let us remind that the results of this section are complemented 
with Appendix A, where we study analytically the supercurrent for the case of 
low transparent interfaces ($r \gg 1$ and $\tau \ll 1$).

\begin{figure}[t]
\begin{center}
\includegraphics[width=0.8\columnwidth,clip]{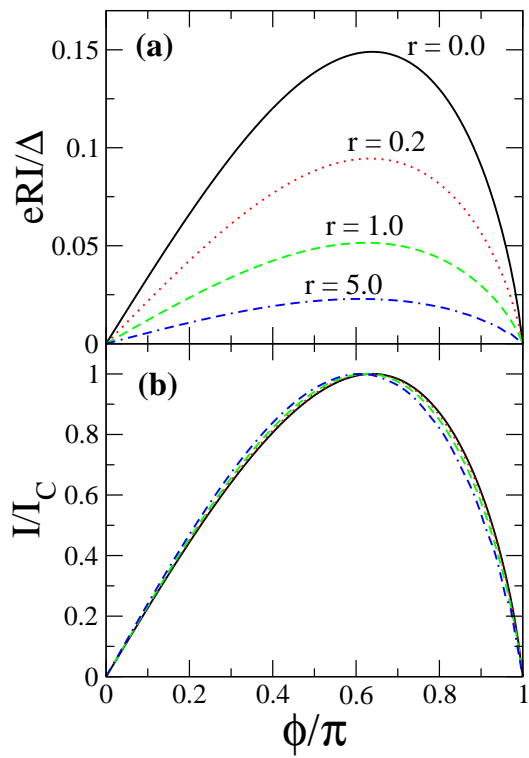}
\caption{\label{superc1} (Color online) Zero-temperature supercurrent-phase
relation of a diffusive SNS junction for $L=8\xi$ ($\Delta = 64 \epsilon_T$)
for different values of the ratio $r=G_{\rm N}/G_{\rm B}$ and $\tau=1.0$. In panel (a) we
show the results for the product $eRI/\Delta$, where $R$ is the normal-state
resistance, while in panel (b) we have normalized the different curves by the 
critical current $I_C$.}
\end{center}
\end{figure}

We start our discussion by analyzing the current-phase relation in the absence
of spin-flip scattering. In Fig.~\ref{superc1} we show this relationship at 
zero temperature for a wire of $L=8\xi$ for different values of the ratio
$r$. As it can be seen, the supercurrent is a non-sinusoidal function of
the phase difference, which reaches its maximum at $\phi \approx 1.27 \pi /2$, 
almost irrespectively of the value of $r$. For the ideal case ($r=0$)
this result agrees with the previous results reported in the 
literature.~\cite{Dubos2001} It is important to stress that in this figure
and in what follows we normalize the current with the total resistance in
the normal state, $R$, which includes the contributions of both the diffusive
wire and the interfaces. For a symmetric junction this resistance can be 
expressed in terms of the ratio $r$ as $R= (1+2r)/G_N$. 

\begin{figure}[t]
\begin{center}
\includegraphics[width=0.8\columnwidth,clip]{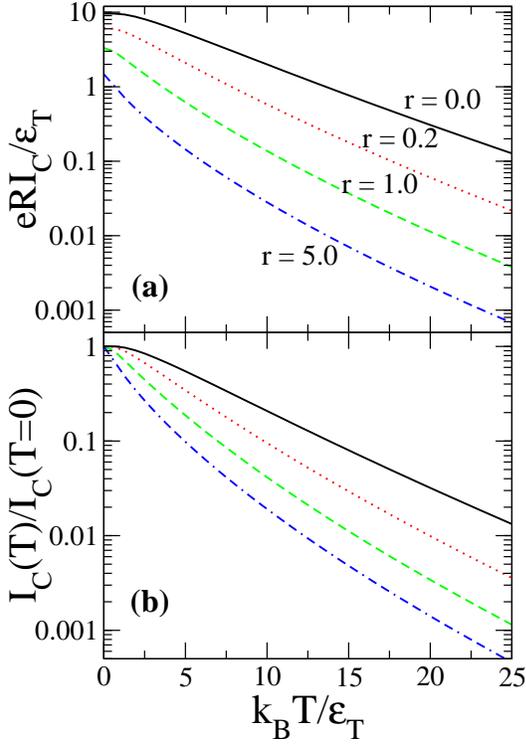}
\caption{\label{superc2} (Color online) (a) Critical current of a diffusive SNS
junction as a function of the temperature for $L=8\xi$ and different values 
of the ratio $r=G_{\rm N}/G_{\rm B}$ and $\tau=1.0$. (b) The same but normalized by the
critical current at zero temperature.}
\end{center}
\end{figure}

Notice that, as one can see in Fig.~\ref{superc1}(b), when the supercurrent is normalized
by the critical current $I_C$, the different results almost collapse into a single
curve. At a first glance, this result seems to suggest that the interface
transparency just enters as a reduction prefactor in the expression of the critical 
current. However, as we discuss in the next paragraph, this is clearly not the 
case at finite temperature.

Let us now turn to the analysis of the temperature dependence of the critical 
current $I_C$. In Fig.~\ref{superc2} we show this dependence for  a wire
of length $L=8\xi$ and different values of $r$. Notice that the temperature is 
normalized with the Thouless energy. The main conclusion that can be extracted
from these results is that the critical current decays faster with temperature
as the interface resistance increases. Moreover, the saturation region at low
temperatures in which the critical current is almost constant shrinks as the
interface resistance increases. For ideal interfaces ($r=0$) this region 
corresponds, roughly speaking, to the range $k_{\rm B} T < \epsilon_T$, while 
for finite $r$ it corresponds to $k_{\rm B} T < \Delta_g$. This illustrates the
fact that is the minigap the scale that controls the magnitude of the supercurrent 
at arbitrary transparency. The faster decay for partially transparent interfaces 
can be confirmed analytically in the limit of very long junctions ($\epsilon_T / 
\Delta \rightarrow 0$). In this case and for perfectly transparent interfaces 
($r=0$), one finds a critical current that decays as $I_C \propto 
(k_{\rm B}T/\epsilon_T)^{3/2} \exp(-L/L_T)$, where $L_T = \sqrt{\hbar D/2 \pi 
k_B T}$ is the thermal length (see Refs.~[\onlinecite{Zaikin1981,Dubos2001}]). 
In the opposite case of opaque interfaces ($r \gg 1$), the result of Appendix 
A indicates that the critical current decays as $I_C \propto (k_{\rm B} T/ 
\epsilon_T)^{1/2} \exp(-L/L_T)$.

\begin{figure}[t]
\begin{center}
\includegraphics[width=0.8\columnwidth,clip]{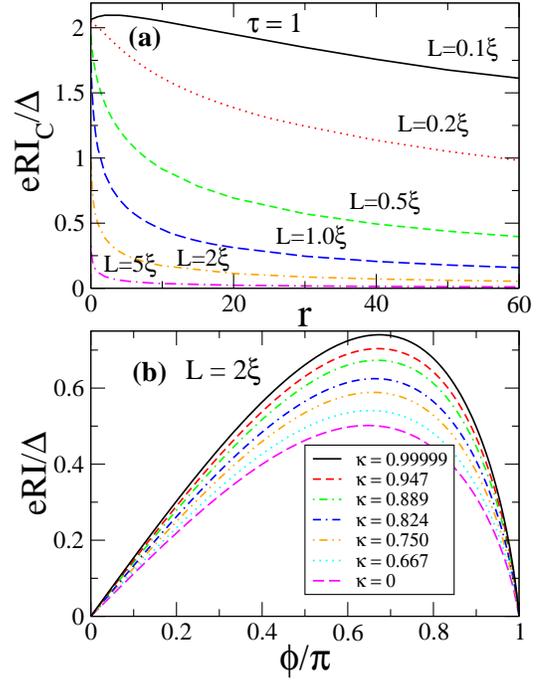}
\caption{\label{superc-r} (Color online) (a) Zero-temperature critical current
of a diffusive SNS system with $\tau=1$ as a function of the parameter $r=G_N / 
G_B$ for different lengths of the normal wire. (b) Current-phase relation at zero 
temperature for an asymmetric junction with $L=2\xi$ and $\tau=1$. The asymmetry 
parameter $\kappa$ is defined as $\kappa = 1 - r_R/r_L$ with $r_R \le r_L$ and 
$r_L+r_R= r_{LR} = \text {const} $. Here $r_{LR}=2$ and $r_L=1,1.5,1.7,1.8,1.9,
1.95,1.99999$}.
\end{center}
\end{figure}                                                                    

The decay of the zero-temperature critical current with the interface resistance 
is examined systematically in Fig.~\ref{superc-r}(a) for different wire lengths 
and fixed transmission $\tau=1$.  After normalizing the curves by the resistance 
in the normal state we find in the limit of very short wires ($L\ll\xi$) a maximal 
critical current at finite $r$ before it slowly decays for large interface 
resistances. Thus, $e R I_{\rm C} \sim \Delta$ in the whole parameter space.  
For wires with $L\geq \xi$ we find a monotonic decay of $e R I_{\rm C} / \Delta$
with increasing $r$. Then, for $\Delta / \epsilon_T \rightarrow \infty$ the energy 
scale of the critical current for large ratios $r$ is determined by an effective 
Thouless energy $\epsilon_{T, \text{eff}} / \epsilon_T \sim A\phantom{.} r^B/(C+r)$.
For instance, when $r\geq 10 $ we can fit the decay of the $eRI_C$ product for the 
special case of a wire with $L=9\xi$ with the help of $e R I_C/ \epsilon_T = 
5.13\phantom{.} r^{0.29}/(0.22+r)$ (see Fig.~\ref{fit}). Here a fitting curve with 
$B=0$ would be proportional to the minigap but would only give a rough estimate of 
$eRI_C(r)$. So far we do not have a good explanation of the factor $r^B$ and the 
numerical value of $B$.

The lower panel of Fig.~\ref{superc-r} shows the current-phase relation at 
zero-temperature for a junction with $L=2\xi$, $\tau=1$ and asymmetric barriers 
as a function of the asymmetry parameter $\kappa=1-r_L/r_R$ that fulfills 
$r_L+r_R=r_{LR}=const$. The critical current shows an enhancement for larger 
asymmetries while the phase difference moves towards $\pi$ as $\kappa$ increases. 
By modeling the diffusive SNS junction as a point contact and averaging the 
current through the different channels over the bimodal distribution for diffusive 
systems, one can understand this trend with the help of the Kirchhoff rules and 
the set of possible shapes of the current-phase relation in this regime.~\cite{Golubov2004}
Furthermore, the formulas of the Appendix A can be generalized to the asymmetric 
case. Then, the $e R I_{\rm C}$ product, Eq.~(\ref{linear-ic}), is proportional to 
$(r_L + r_R)/(r_L r_R)$ what is in agreement with our numerical results.

\begin{figure}[t]
\begin{center}
\includegraphics[width=0.8\columnwidth,clip]{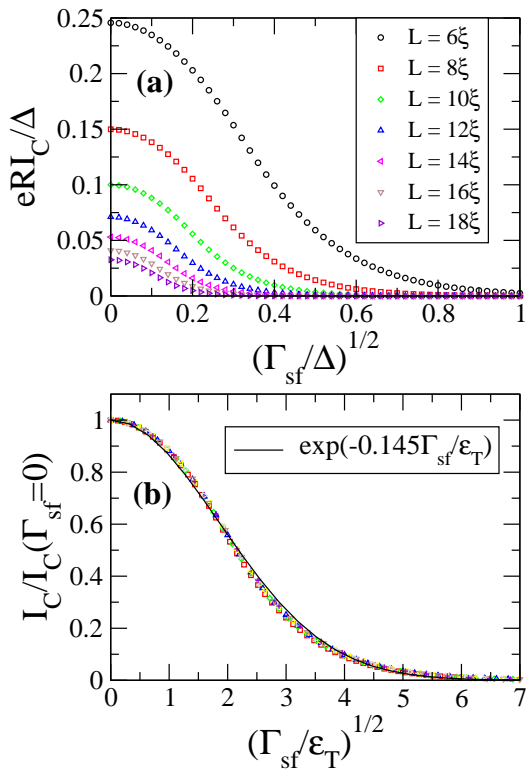}
\caption{\label{superc3} (Color online) (a) Zero-temperature critical current
of a diffusive SNS as a function of the spin-flip rate for different lengths
of the normal wire and ideal interfaces ($r=0$). (b) The same but the critical
current is now normalized by the zero spin-flip rate value and $\Gamma_{sf}$
is normalized by the Thouless energy of the wires. Notice that all the curves
collapse into a single one that can be approximately described by the Gaussian
function $I_C/I_C(\Gamma_{sf} =0) = \exp(-0.145\Gamma_{sf} / \epsilon_T)$
shown as a back solid line.}
\end{center}
\end{figure}

Let us now discuss the influence of a spin-flip mechanism in the supercurrent.
As explained above, the spin-flip scattering may be due to paramagnetic impurities 
and in this case $\Gamma_{sf}$ is proportional to the impurity concentration, or it
may be caused by a magnetic field and in this case $\Gamma_{sf}$ is proportional
to the square of the field. Indeed, this second possibility is much more interesting 
since it offers a natural way to control the strength of the spin-flip scattering 
and, in this sense, it is also more relevant from the experimental point of view.

\begin{figure}[t]
\begin{center}
\includegraphics[width=0.75\columnwidth,clip]{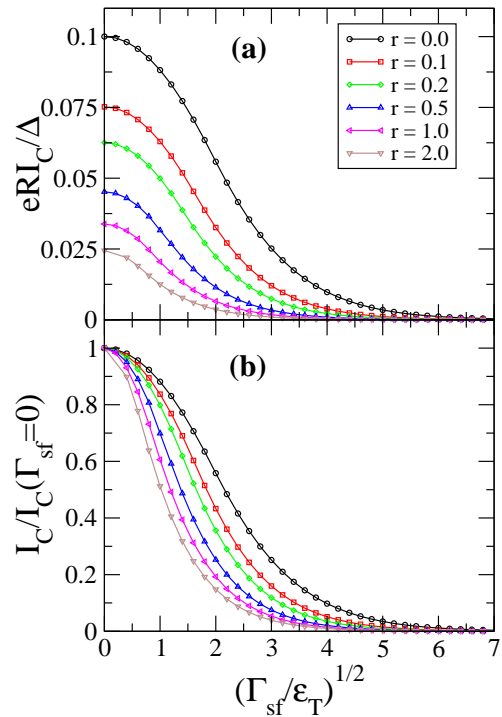}
\caption{\label{superc4} (Color online) (a) Zero-temperature critical current
for a wire length $L=10\xi$ as a function of the spin-flip rate for different
values of the interface resistance and $\tau=0$. (b) The same as in the upper
panel, but the critical current is normalized by its value at $\Gamma_{sf}=0$.
}
\end{center}
\end{figure}

Fig.~\ref{superc3} displays the zero-temperature critical current as a function 
of the spin-flip rate $\Gamma_{sf}$ for different values of the wire length and
ideal interfaces ($r=0$). The reason for plotting the current as a function of
the square root of the rate is that this plot can be seen as the magnetic field 
dependence of the critical current when the normal wire is a thin film. It is 
important to remark that in these calculations we assume that the order parameter
in the leads is not affected by the spin-flip mechanism (such an effect can be 
trivially included). As one can see in Fig.~\ref{superc3} the spin-flip mechanism
causes a decay of the critical current. It is well-known~\cite{Abrikosov1965,Maki1966}
that the spin-flip scattering acts as a pair-breaking mechanism for the Cooper
pairs that penetrate in the normal wire. Such scattering introduces a new relevant
length scale in the problem, namely the spin-flip length $L_{sf} = \sqrt{\hbar D/2 
\Gamma_{sf}}$. When this length becomes smaller than the length of the system and
the thermal length, it dominates the decay of the supercurrent. As we show in 
Fig.~\ref{superc3}(b), when $I_C$ its normalized by its value in the absence
of spin-flip rate, its decay with $\Gamma_{sf}$ becomes universal for relatively 
long wires. Such decay can be phenomenologically fitted with a Gaussian 
function $I_C/I_C(\Gamma_{sf} =0) = \exp(-0.145\Gamma_{sf} / \epsilon_T)$,
as demonstrated in Fig.~\ref{superc3}(b). The analysis detailed in Appendix
A suggests that, in the low transparent regime, the decay follows a law of the
type $I_C \propto (\epsilon_T / 2\Gamma_{sf})^{1/2} \exp(-L/L_{sf})$ at finite
temperature, which numerically is similar to the Gaussian function above.

On the other hand, as one can see in Fig.~\ref{superc3}, there is still a
non-negligible supercurrent even when minigap is completely closed, i.e.
when $\Gamma_{sf} > 5 \epsilon_T$. This phenomenon in the proximity
structure considered here is the equivalent of the well-known gapless
superconductivity in bulk superconductors.~\cite{Abrikosov1965,Maki1966}

In order to understand the role of the interface transparency in the decay
of the critical current as a function of $\Gamma_{sf}$, we present in 
Fig.~\ref{superc4} the results for $I_C$ for a wire of length $L=10\xi$
for different values of the ratio $r$. As it can be seen in particular in
Fig.~\ref{superc4}(b), the critical current decays faster as the interface 
resistance increases. This fact illustrates again that the most relevant 
energy scale at finite transparency is the minigap rather than the Thouless 
energy.

\begin{figure}[t]
\begin{center}
\includegraphics[width=0.75\columnwidth,clip]{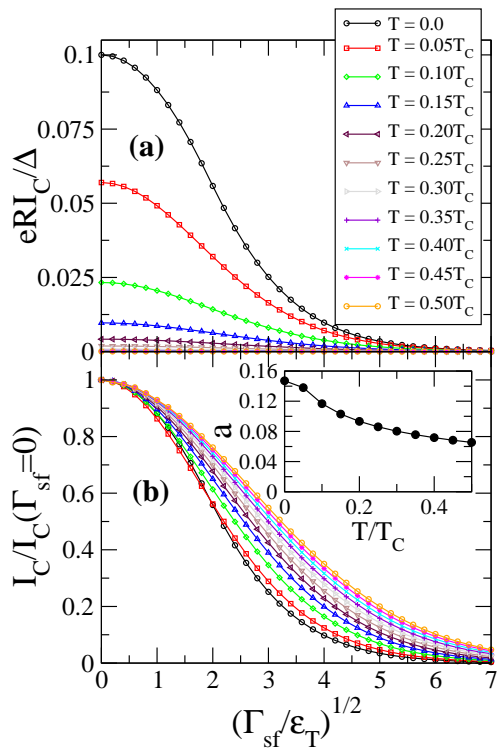}
\caption{\label{superc5} (Color online) (a) Critical current for a wire 
length $L=10\xi$ as a function of the spin-flip rate for different
values of the temperature (in unit of the critical temperature, $T_C$).
(b) The same as in the upper panel, but the critical current is normalized 
by its value at $\Gamma_{sf}=0$. The inset show the temperature dependence
of the constant $a$ used in the Gaussian fits: $I_C/I_C(\Gamma_{sf} =0) = 
\exp(-a \Gamma_{sf} / \epsilon_T)$. }
\end{center}
\end{figure}

Finally, to complete the discussion of the role of the spin-flip scattering,
let us now describe what happens at finite temperatures. In Fig.~\ref{superc5}
one can see the critical current for a wire length $L=10\xi$ as a function
of the rate $\Gamma_{sf}$ for different values of the temperature. The main
conclusion is that by increasing the temperature the decay of the critical
current becomes slower. Such a trend can be understood with the help of the
result of Appendix A.

\section{Conclusions}

With the advances in the fabrication techniques of superconducting hybrid
structures and the development of local measuring probes, it is now 
possible to explore the proximity effect in diffusive metallic nanostructures
in great detail. In this sense, it is highly desirable from the theory side
to elucidate the role of ingredients usually present in experiments such 
as partially transmissive interfaces and pair-breaking mechanisms. With 
this idea in mind, we have presented in this work a detailed analysis of the 
density of states and supercurrent in diffusive SNS junctions. In particular,
we have studied the influence in these two equilibrium properties
of an arbitrary transmission of the interfaces and spin-flip scattering
in the normal wire. Our analysis is based on the quasiclassical theory
for diffusive superconductors (Usadel theory), supplemented by the
boundary conditions put forward by Nazarov.~\cite{Nazarov1999} 

With respect to the local density of states, we have shown that the 
minigap that appears in the normal wire is very sensitive to the 
interface transmission both in the absence and in the presence of a
supercurrent in the system. Moreover, we have shown that the minigap closes 
when the energy rate that describes the spin-flip scattering is a few times 
larger than the minigap in the absence of this type of scattering. This fact 
nicely illustrates that the minigap is indeed the relevant energy scale
for the proximity effect for non-ideal interfaces.

Turning to the analysis of the supercurrent, we have shown that both the magnitude
and temperature dependence of the critical current depend crucially on 
the interface resistance. In particular, the critical current decays faster
with temperature as the interface resistance increases. Moreover, we have
studied how the existence of spin-flip scattering in the normal wire
diminishes the supercurrent and identified the relevant energy and length
scales for its decay. In particular, we have shown that a supercurrent can 
still flow when the minigap is completely closed, which is the analogous
in proximity structures of the well-known gapless superconductivity in bulk 
samples.~\cite{Abrikosov1965,Maki1966} This prediction can be tested
experimentally by using an external magnetic field, as long as the width of
the normal wire is smaller or comparable to the superconducting coherence
length ad it is made of a thin film.~\cite{Belzig1996,Anthore2003}

\acknowledgments

It is a pleasure to acknowledge numerous and fruitful discussions with Sophie 
Gu\'eron, H\'el\`ene Bouchiat, Francesca Chiodi and Meydi Ferrier, and Philippe 
Joyez and H\'el\`ene Le Sueur. We also want to thank them for showing us the 
results of their respective experiments before publication. We also want to 
thank Andrei Zaikin, Gilles Montambaux, Tero Heikkil\"a, Christoph Strunk and
Franziska Rohlfing for useful discussions. F.S.B. acknowledges funding by the Ram\'on y Cajal program.
The work by J.C.H. and W.B. was funded  by the DFG through SFB 513.  
J.C.C. and F.S.B. acknowledge financial support by the  Spanish CYCIT (contract FIS2005-06255). 

\appendix

\section{Linearized equations}

In the limit of very low transparent interfaces ($r \gg 1$ and $\tau \ll 1$), 
the supercurrent can be computed analytically by linearizing the Usadel 
equations.~\cite{Kupriyanov1988} In this appendix we describe how this can 
be done within the formalism presented in section II. 

Assuming that the proximity effect in the normal wire is weak, the coherent
functions are small and the retarded and advanced Green's functions can be 
approximated by (see Eq.~\ref{riccati})
\begin{eqnarray}
\label{l-riccati}
\hat G^{R,A} & \approx & \mp i \pi \left( \begin{array}{cc}
1 & 2 \gamma^{R,A} \\
2 \tilde \gamma^{R,A} & -1
\end{array} \right) .
\end{eqnarray}
Here, the coherent functions $\gamma^R$ and $\tilde \gamma^R$ fulfill the
linearized version of Eqs.~(\ref{g-eq},\ref{gtilde-eq}), which reduce to
\begin{eqnarray}
\label{l-g-eq}
\partial^2_x \gamma^R +  2 \left( \frac{i\epsilon - \Gamma_{sf}}{\epsilon_T}
\right) \gamma^R & = & 0 \\
\partial^2_x \tilde \gamma^R +  2 \left( \frac{i\epsilon - \Gamma_{sf}}{\epsilon_T}
\right) \tilde \gamma^R & = & 0 .
\end{eqnarray}
\noindent
Notice that now the equations for $\gamma^R$ and $\tilde \gamma^R$ are
uncoupled and have an identical form.

The boundary conditions for the previous equations are obtained by linearizing
Eq.~(\ref{g-boundary}) in the following way
\begin{eqnarray}
\label{l-g-boundary}
\mp r \partial_x \gamma^R_{\beta} & = & \frac {\gamma^R_{\alpha}}{1+ \gamma^R_{\alpha}
\tilde \gamma^R_{\alpha}} = -\frac{{\cal F}^R_{\alpha}}{2 \pi i}  \\
\mp r \partial_x \tilde \gamma^R_{\beta} & = & \frac { \tilde \gamma^R_{\alpha}}
{1+ \tilde \gamma^R_{\alpha} \gamma^R_{\alpha}} = -\frac{\tilde 
{\cal F}^R_{\alpha}}{2 \pi i} ,
\end{eqnarray}
where the minus sign is for the left interface and the plus sign for the right one.
Here, ${\cal F}^R_{\alpha}$ and $\tilde {\cal F}^R_{\alpha}$ are the anomalous
Green's functions of the corresponding superconducting lead $\alpha=l,r$.

The solution of Eq.~(\ref{l-g-eq}) with the boundary conditions of 
Eq.~(\ref{l-g-boundary}) can be written as
\begin{equation}
\label{g-linear}
\gamma^R(x) = A^R e^{i \lambda x} + B^R e^{ -i \lambda x} ,
\end{equation}
where $\lambda^2 = 2(i\epsilon - \Gamma_{sf})/\epsilon_T$ and
the constants $A^R$ and $B^R$ can be expressed as
\begin{eqnarray}
\label{constants}
A^R & = & \frac{1}{4 \pi i r \lambda \sin \lambda} \left( {\cal F}^R_r +
{\cal F}^R_l e^{ -i \lambda} \right) \nonumber \\
B^R & = & \frac{1}{4 \pi i r \lambda \sin \lambda} \left( {\cal F}^R_r +
{\cal F}^R_l e^{ i \lambda} \right) \nonumber .
\end{eqnarray}
The solution for the function $\tilde \gamma^R$ is obtained from the solution
for $\gamma^R$ by replacing the functions ${\cal F}^R_{r,l}$ by $\tilde 
{\cal F}^R_{r,l}$.

After linearizing the expression of Eq.~(\ref{superc}), the supercurrent can be written as
\begin{equation}
I = \frac{G_{\rm N}}{e} \int^{\infty}_{-\infty} d\epsilon \; 
\mbox{Re} \left\{ \tilde \gamma^R \partial_x \gamma^R - \gamma^R \partial_x
\tilde \gamma^R \right\} \tanh \left( \frac{\beta \epsilon}{2} \right) . 
\end{equation}

Using the solutions for $\gamma^R$ and $\tilde \gamma^R$, it is straightforward
to show that the supercurrent-phase relation can be written as
\begin{equation}
I = \frac{G_{\rm N}}{e \pi^2 r^2} \sin(\phi) \int^{\infty}_{-\infty} d\epsilon \;
\mbox{Re} \left\{ \frac{-({\cal F}^R_S)^2} { 2 i \lambda \sin \lambda} \right\}
\tanh \left( \frac{\beta \epsilon}{2} \right) ,
\end{equation}
where ${\cal F}^R_S$ is the bulk anomalous Green's function without including the
superconducting phase. This integral can be done analytically and the result
for the critical current is
\begin{equation}
\label{linear-ic}
e R I_C = \frac{4\pi k_{\rm B} T}{r} \sum^{\infty}_{n=0} \frac{ \Delta^2/(\Delta^2 +
\omega^2_n)} { \sqrt{2 \left( \frac{\omega_n + \Gamma_{sf}}{\epsilon_T} \right) }
\sinh \left( \sqrt{2 \left( \frac{\omega_n + \Gamma_{sf}}{\epsilon_T} \right) } 
\right) } ,
\end{equation}
where $\omega_n = (2n+1) \pi k_{\rm B}T$. Here, we have used $R = (1+2r)/G_{\rm N} \approx
2r/G_{\rm N}$. If in particular the temperature is just a few times larger than the
Thouless energy, one just need to keep the first term ($n=0$) in the previous
expression. In the limit of an infinitely long wire this formula reduces
to
\begin{equation}
e R I_C = \frac{4\pi k_{\rm B} T}{r} \left( \frac{\tilde L }{L} \right)
\exp ( - L/ \tilde L) ,
\end{equation}
where the effective length $\tilde L = L_T L_{sf} / \sqrt{L^2_T + L^2_{sf}}$.
Here, $L_T = \sqrt{\hbar D/2 \pi k_{\rm B} T}$ is the thermal length and
$L_{sf} = \sqrt{\hbar D/2 \Gamma_{sf}}$ is the spin-flip length.

\begin{figure}[t]
\begin{center}
\includegraphics[width=0.9\columnwidth,clip]{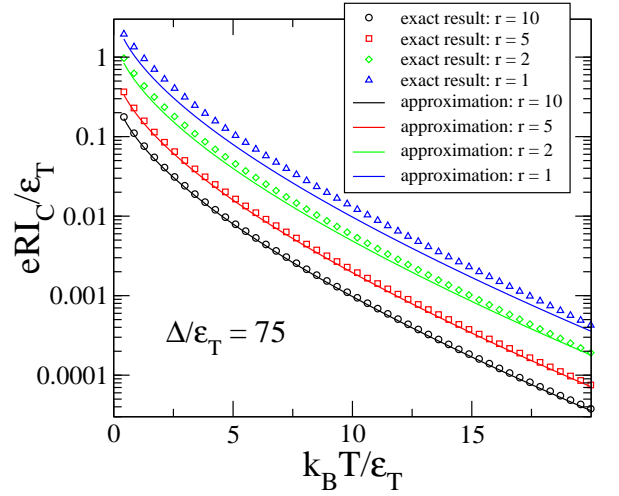}
\caption{\label{approx} (Color online) Comparison between the linearized
result of Eq.~(\ref{linear-ic}) and the exact numerical result for the
temperature dependence of the critical current for a wire with $\epsilon_T 
= 75 \Delta$ in the absence of spin-flip scattering ($\Gamma_{sf}=0$). 
The different curves correspond to different values of the ratio $r = 
G_{\rm N}/G_{\rm B}$ and in the exact result we have used $\tau = 0$. }
\end{center}
\end{figure}

In order to establish the range of validity of the expression of 
Eq.~(\ref{linear-ic}) we have compared this result with the full 
numerical solution of the non-linearized Usadel equations. An example 
of such a comparison for the temperature dependence of the critical current
is presented in Fig.~\ref{approx} for a wire with $\epsilon_T = 75 \Delta$. 
The different curves correspond to different values of the ratio 
$r=G_{\rm N}/G_{\rm B}$ keeping always $\tau=0$, which correspond to a 
tunnel junction. Notice that the approximation of Eq.~(\ref{linear-ic}) 
describes very well the exact results even for values of $r$ very close to 1.

\begin{figure}[b]
\begin{center}
\includegraphics[width=0.85\columnwidth,clip]{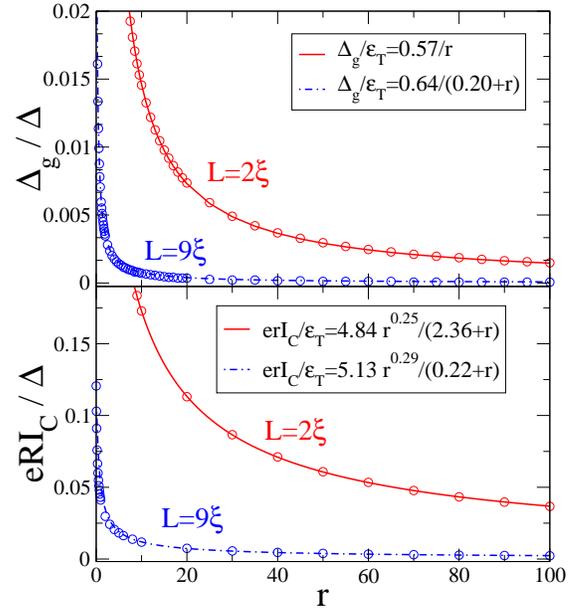}
\caption{\label{fit} (Color online) Minigap and critical supercurrent as
a function of $r$ for the two lengths $L=2\xi$ and $L=9\xi$. The solid lines
show the fitting-curves of the numerically calculated data (circles) in units
of the Thouless energy for $r \gg 1$ .}
\end{center}
\end{figure}

In the limit of highly transparent interfaces, i.e. $r\ll1$, the solution of 
Eqs.~(\ref{g-eq},\ref{gtilde-eq}) demands a more careful treatment.~\cite{Zaikin1981} 
However, in the limit of high temperatures ($k_{{\rm B}}T \gg \Delta$) one can 
obtain an analytical expression for the critical current using the linearized 
solution of Eq.~(\ref{g-linear}) and assuming that this function is continuous 
at the SN interfaces. This means in practice that the constants
$A^R$ and $B^R$ appearing Eq.~(\ref{g-linear}) adopt now the following form
\begin{eqnarray}
A^R & = & \frac{1}{2 i \sin \lambda} \left( \gamma^R_r -
\gamma^R_l e^{ -i \lambda} \right) \nonumber \\
B^R & = & \frac{1}{2 i \sin \lambda} \left( \gamma^R_l e^{ i \lambda} -
\gamma^R_r \right) \nonumber .
\end{eqnarray}
The rest of the calculation is identical and now the result for the critical
current is
\begin{equation}
e R I_{\rm c} =\\
4 \pi k_{\rm B}T \left( \frac{L}{\tilde L} \right) \exp(-L/ \tilde L) .
\end{equation}
In the absence of spin-flip scattering this result reproduces the well-known
result originally derived by Likharev in Ref.~[\onlinecite{Likharev1976}],
which indicates that $I_C \propto (k_{{\rm B}} T/\epsilon_T)^{3/2} \exp(-L/L_T)$. It
has been shown that this dependence describes a broad temperature 
range,~\cite{Zaikin1981,Dubos2001} as long as $k_{{\rm B}}T \gg 
\epsilon_T$.

\section{Numerical fits}

Fig.~\ref{fit} shows some of the numerical
fits mentioned in sections III and IV.



\begin{thebibliography}{}
\bibitem{interaction}
 We assume zero pair interaction in the N region.
\end{thebibliography}
\end{document}